\documentclass[12pt]{iopart}

\usepackage{longtable}
\usepackage{graphicx}

\newcommand{\rbold}{\ensuremath{\textbf{r}}}

\newcommand{\Gr}[1]{\ensuremath{G(#1)}}

\newcommand{\qbold}{\ensuremath{\textbf{Q}}}

\begin{document}
\title{Density profile of water confined in cylindrical pores in MCM-41 silica}
\author{Alan K Soper$^1$}
\address{$ˆ1$ ISIS Facility, STFC Rutherford Appleton Laboratory, Harwell Science and Innovation Campus, Didcot, Oxon, OX11 0QX, UK}
\ead{alan.soper@stfc.ac.uk}

\begin{abstract}
Recently, water absorbed in the porous silica material MCM-41-S15 has been used to demonstrate an apparent fragile to strong dynamical cross-over on cooling below $\sim$220K, and also to claim that the density of confined water reaches a minimum at a temperature around 200K. Both of these behaviours are purported to arise from the crossing of a Widom line above a conjectured liquid-liquid critical point in bulk water. Here it is shown that traditional estimates of the pore diameter in this porous silica material (of order 15\AA) are too small to allow the amount of water that is absorbed by these materials (around 0.5gH$_2$O/g substrate) to occur \textit{only} inside the pore. Either the additional water is absorbed on the surface of the silica particles and outside the pores, or else the pores are larger than the traditional estimates. In addition the low $Q$ Bragg intensities from a sample of MCM-41-S15 porous silica under different dry and wet conditions and with different hydrogen isotopes are simulated using a simple model of the water and silica density profile across the pore. It is found the best agreement of these intensities with experimental data is shown by assuming the much larger pore diameter of 25\AA (radius 12.5\AA). Qualitative agreement is found between these simulated density profiles and those found in recent empirical potential structure refinement (EPSR) simulations of the same data, even though the latter analysis did not specifically include the Bragg peaks in the atomistic structure refinement. It is shown that the change in the (100) peak intensity on cooling from 300K to 210K, which previously has been ascribed to a change in density of the confined water on cooling, can equally be ascribed to a change in density profile at constant average density. It is further pointed out that, independent of whether the pore diameter really is as large as 25\AA\ or whether a significant amount of water is absorbed outside the pore, the earlier reports of a dynamic cross-over in supercooled confined water could in fact be a crystallisation transition in the larger pore or surface water. 
\end{abstract}
\submitto{\JPCM}
\maketitle

\section{Introduction}
\label{intro}
In a series of papers which extend back at least 8 years, S-H Chen and colleagues have performed detailed studies of the structural and dynamical properties of confined water.\cite{savedrecs2004:1, savedrecs2005:2, savedrecs2005:1, savedrecs2006:5, savedrecs2006:4, savedrecs2006:3, savedrecs2006:2, savedrecs2006:1, kumar2006:1, savedrecs2007:1, savedrecs2008:4, savedrecs2008:3, savedrecs2008:2, savedrecs2008:1, kumar2008:1, savedrecs2009:2, savedrecs2009:1, savedrecs2010:4, savedrecs2010:3, savedrecs2010:2, savedrecs2010:1} The methods used include quasi-elastic neutron scattering (QENS), small angle scattering (SANS), NMR, and computer simulation, and there is a general finding that confined water appears to undergo an dynamical crossover from Arrhenius to Vogel-Fulcher (fragile to strong, FS) behaviour at a temperature near 220K. The hypothesis is made that this FS transition coincides with the crossing of a Widom line in the phase diagram of supercooled water which extends above a purported liquid-liquid critical point \cite{savedrecs2005:1}. The same hypothesis is made concerning the apparent observation of a density minimum in confined water at a lower temperature near 200K \cite{savedrecs-32007:1}, with the lowest density occurring at temperatures below that of the Widom line, and the maximum decrease in density occurring at the Widom line itself. However the dynamic crossover phenomenon is not quite a universal phenomenon of confined water since in some cases, where the surface is hydrophobic, the effect appears to be absent or shifted to a much lower temperature regime,\cite{savedrecs2009:3,savedrecs2009:4} suggesting it is impacted to some extent by the nature of the surface itself.

Others have questioned the interpretation of the dynamic data in terms of a dynamic crossover. Both Cerveny \textit{et. al.} \cite{waterrefs2006:3} and Swenson \cite{swenson2006:8} issued comments on Liu \textit{et. al.}\cite{savedrecs2005:1}. Cerveny \textit{et. al.} argue that the observed fragile-to-strong crossover is in fact due to the onset of confinement effects and quote several related cases of confined water where the same behaviour in the QENS data has been seen, and also quote the case of polymer blends where the same trend is seen. Swenson argues that if the QENS data are taken literally they would extrapolate to a glass transition temperature of 50K, which would be unacceptably low. Like Cerveny \textit{et. al.} he also argues that what the QENS data are seeing is the effect of confinement killing the $\alpha$ relaxation process in water, rather than any FS transition. Subsequently Swenson and coworkers have published a dielectric relaxation of study of water highly confined in MCM-41 \cite{swenson2007:1}. They observe no obvious transition from Arrhenius to Vogel-Fulcher-Tammann (VFT) behaviour in the dielectric relaxation time of the water as a function of temperature, particularly at the temperature where this transition occurs in the QENS and NMR data. In a more recent study of water near a protein surface Doster \textit{et al.} also suggest an alternative explanation based on a glass transition scenario, instead of the purported fragile-strong crossover.\cite{doster2010:1} Even more recently Limmer and Chandler have questioned whether atomistic computer simulations of low temperature water are capable of observing a liquid-liquid transition, arguing instead that what is observed in the simulations is a non-equilibriated liquid-crystal transition \cite{2011arXiv1107.0337L}. 

To try to understand what might be going on with the structure of the water confined in a pore of MCM-41-S15 Mancinelli \textit{et al.} undertook a combined neutron scattering and computer modelling study of MCM-41-S15 silica both dry and with absorbed light and heavy water, the latter samples giving a marked scattering contrast due the change in neutron scattering amplitude between heavy and light water.\cite{manc2009:1,manc2010:1,manc2010:2} According to Liu \textit{et al.}\cite{savedrecs-32007:1} the diameter of the pores in this material is 15$\pm$1\AA. Two factors stand out from that study, namely, (a) the density profile of water in the silica pore is apparently not uniform across the pore, and (b) the density profile changes markedly when the temperature is lowered through 210K, being most uniform at that temperature and less uniform at temperatures either side of this value. The point here is that the measurement of a density minimum in confined water was achieved by measuring the intensity of the (100) Bragg peak in MCM-41-S15 silica as a function of temperature.\cite{savedrecs-32007:1} However the intensity of that peak will depend not only on the overall density of water in the pore but also on the way that water is distributed across the pore. Hence a change in intensity of the Bragg peak could equally be regarded as a due to a change in the overall density, or to a change in the distribution of density across the pore: this peak by itself cannot distinguish between these two scenarios.

In this paper I show that the amount of water absorbed by MCM-41-S15, typically about 0.5gH$_2$O/g substrate, is too large to be filling only the pores, assuming their nominal diameter of 15\AA. This means that either their actual diameter for absorbing water is larger than 15\AA\ or else some of the absorbed water appears on the surface of the silica particle and not in the pore itself. Either way, the discrepancy in pore sizes casts some doubt on the earlier observations of a fragile to strong dynamical transition in supercooled water, since water in the larger pore, or alternatively water on the surface of the silica particles, is likely to freeze on cooling, rather than go through a dynamical transition. Such surface water would of course, if present, be just as visible in scattering experiments as water in the pore.

In addition I assess the intensity of the Bragg peaks in MCM-41-S15 silica as a function of water content (dry or fully hydrated) and of whether the water is protiated or deuteriated. As will be seen the availability of scattering data on each of these cases, namely dry, fully hydrated with H$_2$O, and fully hydrated with D$_2$O provides a significant constraint on the shape of the density profile. In particular it is shown that marked changes in the amplitude of these peaks, particularly for the deuterated material, will occur if the density profile of the water against the silica substrate changes, even if the overall mass of water present is unchanged. Hence the use of the amplitude of these peaks to assess the density of the confined water \cite{savedrecs-32007:1} may be unreliable. More importantly it is suggested that this analysis could be applied to other fluids absorbed on MCM-41-S15 silica and used to assess the likely surface density profile in those systems as well.

\section{\label{theory}Theory of Bragg scattering for a hexagonal lattice}

\begin{figure}[htp]
\centering
\includegraphics[width=1.0\textwidth]{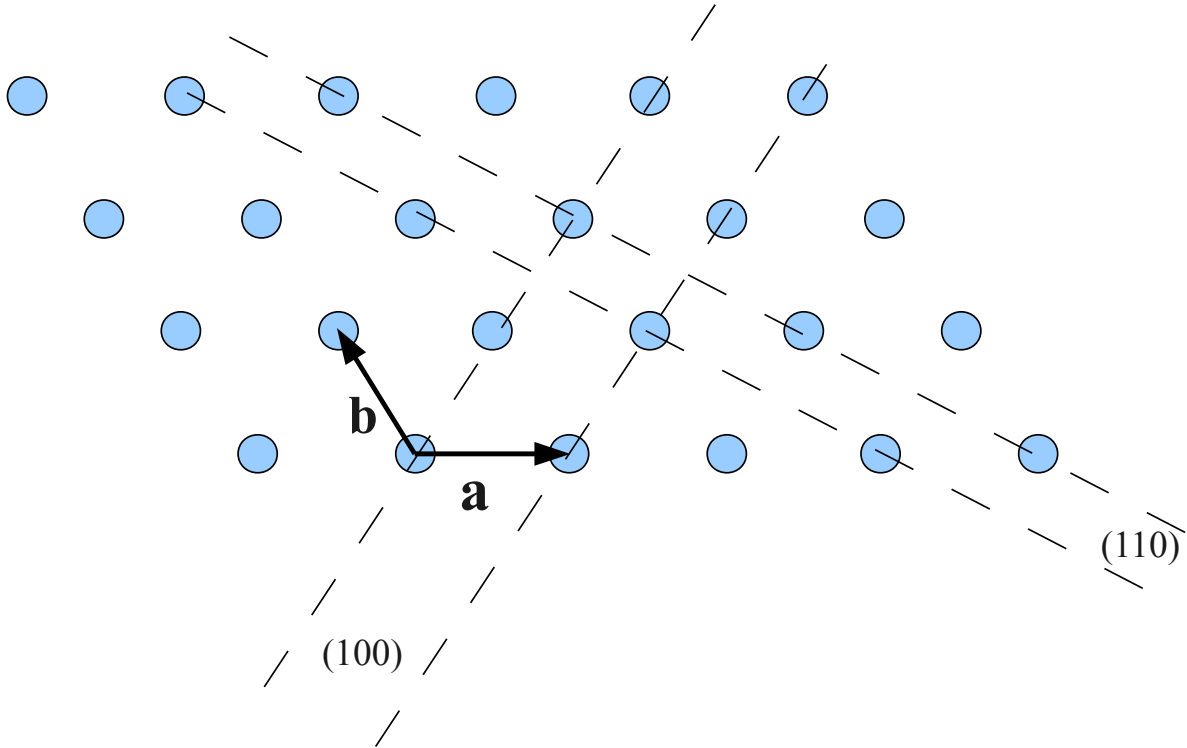}
\caption{Schematic of hexagonal lattice of cylinders as found in MCM-41 silica. The lattice of hollow cylinders (shaded blue) is formed in an amorphous silica matrix. The axis of the cylinders is at right angles to the page. The first two Bragg planes, (100) and (110), are shown as pairs of dashed lines. The cylinders can be filled with various liquids, including water as in the present instance. This drawing is not necessarily to scale.}
\label{fig-hexagonallattice} 
\end{figure}

Figure \ref{fig-hexagonallattice} shows the first two atomic planes in the hexagonal lattice which give rise to the (100) and (110) Bragg reflections in $Q$ (reciprocal) space. It is assumed here the cylinders extend indefinitely along their axes (perpendicular to the diagram) although in practice they are around 1$\mu$m long. Provided this linear extent is much larger than the radius, as in the present case, the finite length of the real cylinder will not significantly affect the outcome of the present analysis.

\begin{figure}[htp]
\centering
\includegraphics[width=1.0\textwidth]{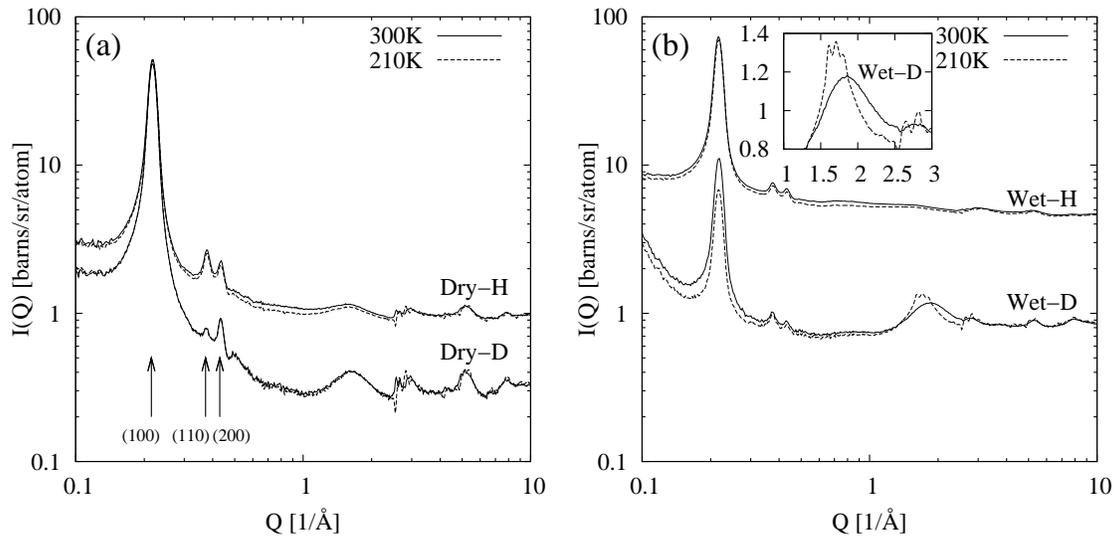}
\caption{Differential scattering cross section for dry (a) and wet (b) MCM-41 silica showing the low $Q$ Bragg peaks and the wider $Q$ scattering used to estimate the atom-scale structure.\cite{manc2009:1} Solid lines correspond to the 300K data, while the dashed lines correspond to 210K data. Note the logarithmic intensity scale. The inset in (b) shows the main water diffraction peak for the Wet-D sample at the same two temperatures. Note the marked shift of this peak to lower Q at 210K compared to 300K and the appearance of small Bragg-like features at 210K.}
\label{fig-mdcs} 
\end{figure}

Figure \ref{fig-mdcs} shows the actual diffraction data from the dry and wet MCM-41 silica and the Bragg peaks referred to above. If $d$ is the spacing between cylinders, then spacing between the planes is $d_{100}=\frac{\sqrt{3}}{2}d$ and $d_{110}=\frac{d}{2}$. Since the reciprocal lattice vector for these reflections is perpendicular to the crystallographic \textbf{c} axis (which is parallel to the cylinder axis), these reflections can tell us nothing about the distribution of density along the cylinder axis. The Fourier transform of an infinitely long, infinitely thin cylinder will be a sheet of intensity in reciprocal space which lies normal to the axis cylinder. However the material in this case is polycrystalline with an assumed random orientation of crystallites, so the scattering intensity has to be averaged over these orientations.The Bragg peaks have to be further convoluted with the instrumental resolution function, which tends to be broad at low scattering angles. The $Q$ values of these reflections will be $\frac{2\pi}{d_{hkl}}$ which works out, assuming $d=33.1$\AA, to be 0.219\AA$^{-1}$ for (100), 0.379\AA$^{-1}$ for (110), and 0.438\AA$^{-1}$ for (200). These values are close to the observed peak positions. 

In practice the cylinders are not infinitely thin and will possibly have density inhomogeneities both along and perpendicular to the cylinder axis. Hence the Bragg intensities will be modified by the form factor arising from the finite size of, and density variation within, the cylinders. For simplicity we assume the scattering length density inhomogeneities are axially symmetric and so are a function of displacement from the cylinder axis (perpendicular radius, $\rbold_{perp}=\rbold_x+\rbold_y$) and distance along the cylinder ($z$) only. Hence we can represent these inhomogeneities as $\rho(x,y,z)\equiv\rho(r_{perp},z)$. The scattering amplitude of the cylinder in reciprocal space is then written:

\begin{equation}
C(Q)=\int dxdydz \rho(x,y,z)\exp\left(i\qbold\cdot\rbold\right)
\label{cylinderamplitude1}
\end{equation}								
where $\rbold=\rbold_x+\rbold_y+\rbold_z=\rbold_{perp}+\rbold_z$. $\qbold$ can also be split into perpendicular and parallel components: $\qbold=\qbold_{perp}+\qbold_{z}$. Hence $\qbold\cdot\rbold=Q_{perp}r_{perp}\cos\alpha+Q_{z}z$ and $dxdy=r_{perp}dr_{perp}d\alpha$, where $\alpha$ is the angle between $\qbold_{perp }$ and $\rbold_{perp}$. For the (100), (110) and (200) reflections, $Q_{z}z$ is zero by construction, so the integral over $z$ can  be performed directly and there can be no sensitivity in the Bragg peaks to the dependence of $\rho\left(r_{perp},z\right)$ on $z$. Hence we replace $\rho\left(r_{perp},z\right)$ by $\overline{\rho}\left(r_{perp}\right)$, which represents the density as a function of perpendicular radius across the pore and averaged over the length of the pore. This is the radial density function calculated in the Empirical Potential Structure Refinement (EPSR) simulations \cite{manc2009:1}, and will be referred to here as the density profile across the pore.

Of course the Bragg intensity itself will be related to the value of $\vert \overline{C}(Q_{(hkl)})\vert^{2}$, and to calculate this intensity the density profile across the pore has be convoluted with the hexagonal distribution of pores and averaged over the orientations of the crystallites. The simplest way to achieve this \cite{manc2009:1} is to set up a scattering length density function, $n(\rbold)= \langle b(\rbold)\rangle\rho(\rbold)$, where $\langle b(\rbold)\rangle$ is the average scattering amplitude and $\rho(\rbold)$ is the local atomic number density at position $\rbold$ in the substrate or pore, and then perform the autocorrelation of $n(\rbold)$ to form a scattering cross section radial distribution function, 
\begin{equation}
\Gr{r}=\left\langle\int_{V}n(\rbold^{\prime})n(\rbold-\rbold^{\prime})d\rbold\right\rangle,
\label{eq-gofr}
\end{equation}
 where the angle brackets correspond to the average over the orientation of $\rbold$. The scattering cross section is then calculated by Fourier transforming this to $Q$ space:

\begin{equation}
I(Q)=\frac{4\pi}{Q}\int r\Gr{r}\sin Qr dr
\label{eq-iofq}
\end{equation}
This is the procedure adopted in the present work. For pure bulk silica (SiO$_2$), $\langle b(\rbold)\rangle = 5.25$fm/atom, for H$_2$O $\langle b(\rbold)\rangle = -0.56$fm/atom, and for D$_2$O $\langle b(\rbold)\rangle = 6.38$fm/atom. Assumed average atomic number densities will be 0.0663 atoms/\AA$^{3}$ and 0.086 atoms/\AA$^3$ for silica and water respectively. Silanol groups are attached to the inside of the pore, and these will be treated as OH groups ($\langle b(\rbold)\rangle = 1.03$fm/atom for OH and $\langle b(\rbold)\rangle = 6.24$fm/atom for OD). The fraction of Si atoms that have an OH attached is believed to be around 0.2 and is determined by inspection of the data as discussed in \cite{manc2009:1} and below.

\section{\label{experiment}Experiment}

This paper will focus only on the intensities of the (100),(110) and (200) Bragg peaks for the wet and dry MCM-41-S15 using both H and D substituted water and for the two temperatures 300K and 210K. The experiment itself has been fully described in the preceding papers, \cite{manc2009:1,manc2010:1,manc2010:2} and involves measuring the wide angle neutron diffraction pattern, over a wide Q range of 0.07 to 50\AA$^{-1}$, from a series of dry and wet porous silicas, MCM-41-S15, whose nominal pore radius is 15\AA. The wet and dry samples are measured alternatively with H then deuterium (D) substitution on the water and silanol groups. In order to get the height of the Bragg peak onto a common scale for different water contents and different temperatures, the measured diffraction data are here renormalised onto a scale of differential cross section per scattering unit, with the scattering unit defined by the dry MCM-41 samples, namely $c_{Si}=0.28$, $c_{O}=0.61$ and $c_{H}=0.11$, where $c_{\alpha}$ is the atomic fraction of component $\alpha$. (See Table 1 in \cite{manc2009:1}.) The choice of scattering unit to normalise to is not particularly important for the present study. What is important is that it is the same scattering unit for all samples, H or D, wet or dry. That way the relative heights of the scattering levels and Bragg peaks can be compared between wet and dry samples.

The renormalised data are shown in Fig. \ref{fig-mdcs}. Several features are notable from these graphs. Firstly, the high $Q$ scattering level is determined by the composition of each sample. Since all the data have been normalised per atom of substrate, as the composition changes, so the high $Q$ scattering level changes. There is a substantial change in the incoherent neutron scattering level between H and D isotopes (6.53 barns/sr/atom for H, 0.61 barns/sr/atom for D, 1 barn = $10^{-28}$m$^2$), and as a result all the samples containing H have a significantly higher high $Q$ scattering level compared to those with D. The H component in the dry samples arises from silanol groups on the pore surfaces. In fact the change in this scattering level between H and D isotopes in the dry samples is consistent with an assumed silanol fraction of $\approx$0.2, as discussed in \cite{manc2009:1}. The change in the scattering level between dry and wet samples is consistent with a water adsorption of approximately 0.3g water/g substrate for both H and D samples. This is somewhat lower than the 0.43g/g stated in \cite{manc2009:1,manc2010:2}, but as will be seen below a discrepancy can arise because of the assumed radius of the pore, and the different numbers quoted here and in \cite{manc2009:1} may not be outside the experimental uncertainties in determining this value.

A second observation is that in the dry samples, the main (100) Bragg peak barely changes in amplitude between the H and D samples, while the same peak in the wet samples changes drastically, with intensity in the D sample dropping by a factor of about 7 compared to the corresponding H sample at the same hydration level. This large drop in the wet-D sample arises because the difference in scattering length density between pore region and silica region is large for the H sample, $\approx -0.3$fm/\AA$^{3}$, but smaller, $\approx 0.2$fm/\AA$^{3}$, for the D sample (and the Bragg intensity will of course be proportional to the square of this amplitude). However the difference in scattering length density is not the only factor that contributes to the height of the Bragg peak, as will emmerge shortly. Table \ref{tab-relinten} lists the relative intensities of the (100) peak as a function of hydration, isotope and temperature, after substracting the underlying background from interfacial scattering, as discussed in \cite{savedrecs-32007:1}.

\begin{table}
\caption{\label{tab-relinten} Relative heights of the (100) Bragg peak for MCM-41-S15 for dry and wet hydrations and at 300K and 210K as a function of hydrogen isotope, H or D. All intensities are normalised to the intensity of the dry-D sample.\hspace*{12pt}}
\begin{center}
\begin{tabular}{| l | c | c |}
\hline
Sample & 300K & 210K \\
\hline
Dry-D  & 1.00 & 1.00 \\
Dry-H  & 1.03 & 1.03 \\
Wet-D  & 0.21 & 0.12 \\
Wet-H  & 1.40 & 1.32 \\
\hline
\end{tabular}
\end{center}
\end{table}

At the same time the (110) and (200) peaks change with the different hydration levels and isotope substitutions. In particular the intensities of these peaks is markedly weaker for the D samples compared to H, particularly for the dry-D samples. (Note the logarithmic intensity scales in Figure \ref{fig-mdcs} when comparing different data.)

Hence these three aspects of the Bragg diffraction peaks as a function of hydration and isotope need to be reproduced by any model of the water density profile.

\section{\label{dataanalysis}Data analysis}

\subsection{\label{rhomodel}Modelling the density profile}

For the purpose of the present discussion the water density profile is divided into three regions, namely a `core' region, $0 \leq r \leq r_{C}$, representing water away from the silica interface, an `interfacial' region, $r_{C} \leq r \leq r_{I}$, and an `overlap' region, $r_{I} \leq r \leq r_{P}$ where water is interacting with the OH groups on the silica surface. The water density is specified separately for each of these regions, but is otherwise assumed uniform within in each region:

\begin{equation}
\bar{\rho}_{W}(r) = \cases{
\rho_{C}, & for $0 \leq r < r_{C} $\\
\rho_{I}, & for $r_{C} \leq r < r_{I}$ \\ 
\rho_{O}, & for $r_{I} \leq r < r_{P}$ \\
0       , & for $r_{P} < r$ \\
} 
\label{eq-rhowater}
\end{equation}
The model therefore contains no atomistic information and cannot be used to calculate the structure functions at larger $Q$ values. This choice of three water regions is intended as a simplistic representation of the actual density profile found in \cite{manc2009:1,manc2010:2}. The choice is certainly not unique, but is probably the minimum required to represent the actual density distribution obtained in \cite{manc2009:1}.

The silica density profile is assumed to be zero in the core and interface regions, but can have a finite value in the overlap region, where it is treated as consisting entirely of OH groups (with the stoichiometry O:H in the ratio of 1:2 to ensure charge neutrality). No Si is allowed in this region. This therefore assumes half of the oxygen atoms of OH groups do not occur in the overlap region but in the bulk region instead. The remainder of the volume is assumed to be the density and composition of pure bulk silica (SiO$_2$). The density of the OH groups in the overlap region is obtained from the specified fraction of silanol groups (Si(OH)$_2$) in the substrate as a whole, namely $\approx$ 0.2, and with the assumption that no H atoms occur in the bulk silica region, and no Si atoms occur in the overlap region:

\begin{equation}
\bar{\rho}_{S}(r) = \cases{
0, & for $0 \leq r < r_{C} $\\
0, & for $r_{C} \leq r < r_{I}$ \\ 
\rho_{OH}, & for $r_{I} \leq r < r_{P}$ \\
\rho_{B} , & for $r_{P} < r$ \\
} 
\label{eq-rhosubstrate}
\end{equation}
where $\rho_{B}$ = 0.0663 atoms/\AA$^3$, as stated above.

The above definitions also enable the respective volumes per unit length along the pore axis to be defined:
\begin{eqnarray}
V_C &=& \pi r_C^2 \label{eq-vc}\\
V_I &=& \pi (r_I^2-r_C^2)  \label{eq-vi}\\
V_O &=& \pi (r_P^2-r_I^2)  \label{eq-vo}\\
V_B &=& V_{total}- \pi r_P^2
\end{eqnarray}
where $V_{total}=\frac{\sqrt{3}}{2} d^2$ is the total volume of the hexagonal unit cell per unit length. 

If $x$ is the silanol fraction then the average atomic fractions of Si, O and H in the substrate are given by $c_{Si}=1/(3+3x)\mathrm{,\ } c_{O}=(2+x)/(3+3x) \mathrm{, and\ }c_H=2x/(3+3x)$\cite{manc2009:1}. If $c_{Si,B}=1/3$ is the atomic fraction of Si in the bulk region and $c_{Si,O}=0$ is the assumed atomic fraction of Si in the overlap region, then the atomic density of the substrate in the overlap region can be calculated from:

\begin{equation}
\rho_{OH}=\frac{(c_{Si,B}-c_{Si})\rho_{B}V_B}{(c_{Si}-c_{Si,O})V_O}
\label{eq-rhoshelf}
\end{equation}
while the average density of the substrate is given by $\bar{\rho_S}=\left(V_O\rho_{OH}+V_B\rho_B\right)/\left(V_O+V_B\right)$. Note that setting the silanol fraction to zero in this model is equivalent to setting the volume of the overlap region to zero, since then $c_{Si,B}=c_{Si} $ in (\ref{eq-rhoshelf}) and there is then no substrate density in the overlap region. In the analysis that follows the thickness of the overlap region, as determined by the difference $r_P-r_I$ was treated as an adjustable parameter to give agreement Bragg peak intensities with experiment, while the silanol fraction is held constant at 0.2, in accord with the earlier experiment \cite{manc2009:1}.

The outer radius of the overlap region, $r_P$, is used to determine the total number of atoms of water in the pore, $N_w=\langle{\rho}_W\rangle V_w$, where $V_w=V_C+V_I+V_O=\pi r_P^2$ per unit distance along the pore.

In order to evaluate the radial density function, $G(r)$ and hence the scattering intensity, $I(Q)$, a hexagonal supercell is constructed consisting of 21 unit cells along each of the \textbf{a} and \textbf{b} crystallographic axes, with a cylindrical pore at the centre of each unit cell, and with an equivalent extension along the \textbf{c} axis which is at right angles to the hexagonal plane and parallel to the pore cylinder axes. Points are chosen at random within this supercell and assigned weights according to the substrate and/or water scattering densities at the  position defined by the point, i.e. in the core, interfacial, overlap or bulk regions. The auto correlation function was then calculated directly from these weighted points out to a distance of $r=\frac{\sqrt{3}}{2}10d$, and, by using a standard Fourier transform, the corresponding scattering intensity in $Q$ space is computed. Note that this simulation is not an atomistic empirical potential structure refinement (EPSR) simulation as performed by Mancinelli \textit{et al.} \cite{manc2009:1} and so cannot reproduce details of the atomic and molecular structure. 

\subsection{\label{poreradius}Estimating the pore radius}
								
\begin{figure}[htp]
\centering
\includegraphics[width=1.0\textwidth]{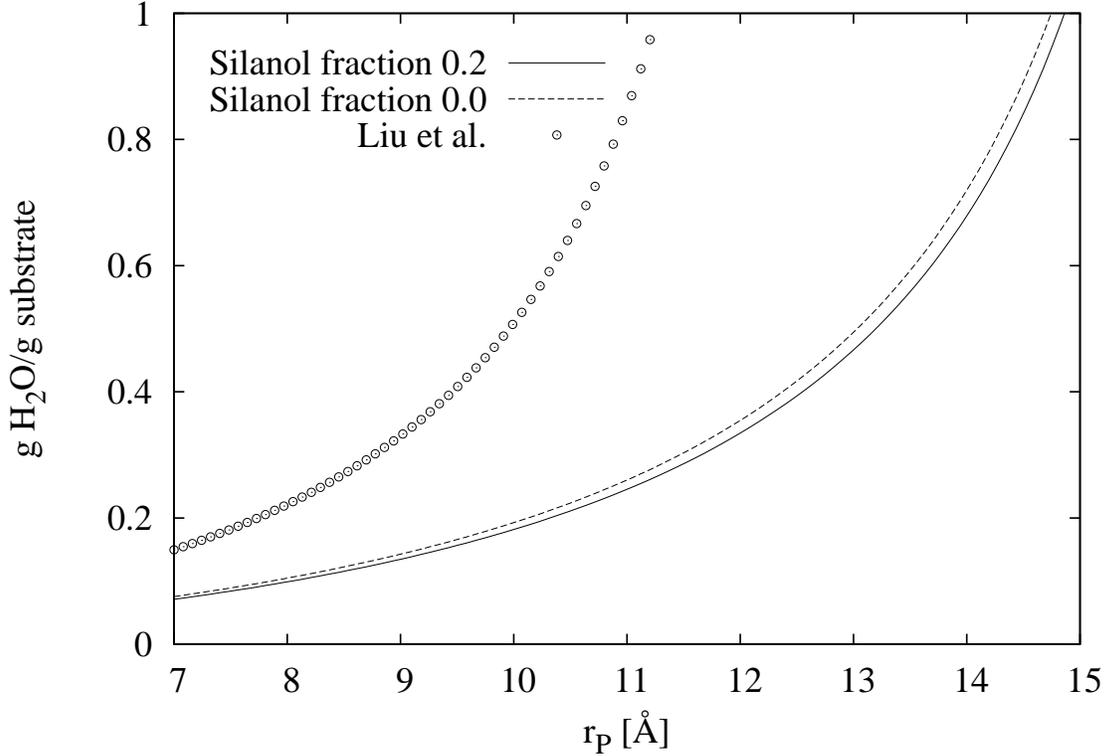}
\caption{Calculated mass of water per mass of substrate as a function of pore radius, for two silanol fractions, 0.0 (dashed) and 0.2 (solid). For the latter case it was assumed the inner radius of the overlap region, $r_I$ extended 3\AA\ below the pore radius, $r_P$. Note that a mass adsorption of $\approx$ 0.4g/g can only be achieved if the pore radius is of order 12.5\AA\ or greater. This calculation assumes the average water atomic number density is $\langle{\rho}_W\rangle=0.085$atoms/\AA$^3$ as stated in \cite{manc2009:1}. Also shown is the case of Liu \textit{et al.} \cite{savedrecs-32007:1} assuming zero silanol fraction and a pore separation of 25.3\AA\ (circles). See text for more discussion.}
\label{fig-massadsorbed} 
\end{figure}

The definitions of the previous section allow us to estimate the mass of water that can theoretically be adsorbed by different pore radii. If $A_W=6$ is the average atomic mass per atom of water, $A_S=\sum_{\alpha}c_{\alpha}A_{\alpha}$ is the average atomic mass per atom of substrate, and we assume water occupies the pore uniformly, then the mass of water adsorbed per mass of substrate is given by 

\begin{equation}
M\left(r_P\right)= \langle{\rho}_W\rangle \pi r_P^2A_W/\left(V_{total}- \pi r_I^2\right)\bar{\rho_S}A_S.
\label{eq-massrp}
\end{equation}
This function is plotted in Figure \ref{fig-massadsorbed} for the case where the overlap thickness is held constant at 3.0\AA\ and the pore radius, $r_P$, varied. Note that, as stated above, if the silanol fraction is set to zero (dashed line in Fig. \ref{fig-massadsorbed}) this is equivalent to having zero substrate density in the overlap region. This figure demonstrates that neither the assumed overlap thickness nor silanol fraction makes a major difference to the conclusions about the possible pore radii for a given water mass adsorption.

It will be noted immediately that if, according previous work \cite{manc2009:1} the mass of water adsorbed is of order 0.4g H$_2$O/g substrate, the radius of the pore must be of order 12.5\AA\ (diameter $\approx$ 25\AA) - given the pore spacing of 33.1\AA\ there would not be enough volume for so much water in pores of significantly smaller radius. Moreover the composition of the substrate, i.e. whether silanols are assumed to be present or not, and whether or not there is an overlap region, does not affect this conclusion very significantly. 

One notable discrepancy between the data of Mancinelli \textit{et al.} \cite{manc2009:1} and that of Liu \textit{et al.} \cite{savedrecs-32007:1} is that for the latter case the (100) peak occurs at $Q=0.287$\AA$^{-1}$, not $Q=0.219$\AA$^{-1}$ as here, even though the samples of MCM-41 are nominally the same. Such a large discrepancy in peak positions for the same material is surprising, but for the latter case would imply a pore separation of 25.3\AA\ which is much smaller than the 33.1\AA\ pore separation in the present sample. Therefore the mass ratio for this smaller pore spacing is also shown in Fig. \ref{fig-massadsorbed} as the dotted line. In that work the stated pore diameter is 15$\pm$1\AA\, meaning a pore radius of 7.5\AA, while the stated water adsorption is 0.5g D$_2$O/g substrate, which is equivalent to 0.45g H$_2$O/g substrate. However from Fig. \ref{fig-massadsorbed} it is clear that such a large water absorption \textit{inside} the pore is impossible for the specified pore radius of 7.5\AA, for which the maximum amount of water that can be absorbed is $<0.2$gH$_2$O/g substrate. A similar comment applies to other work, \cite{savedrecs2005:1}, where the pore diameter is stated to be 14\AA\ and the stated water absorption is 0.5gH$_2$O/g substrate. 

There are two possible explanations of these discrepancies, namely either the pore radii are larger than is stated in any of the referenced papers, or else a significant amount of the absorbed water does not enter the cylindrical pores but is absorbed on the surface of the silica particles. Such surface absorbed water will most likely behave differently on cooling compared to the pore water, but will be equally visible in scattering experiments, and so can affect both the observed dynamic and Bragg diffraction behaviour. Hence it becomes rather important to establish the true size of the pores and the amount of surface water present. 

\begin{figure}[htp]
\centering
\includegraphics[width=1.0\textwidth]{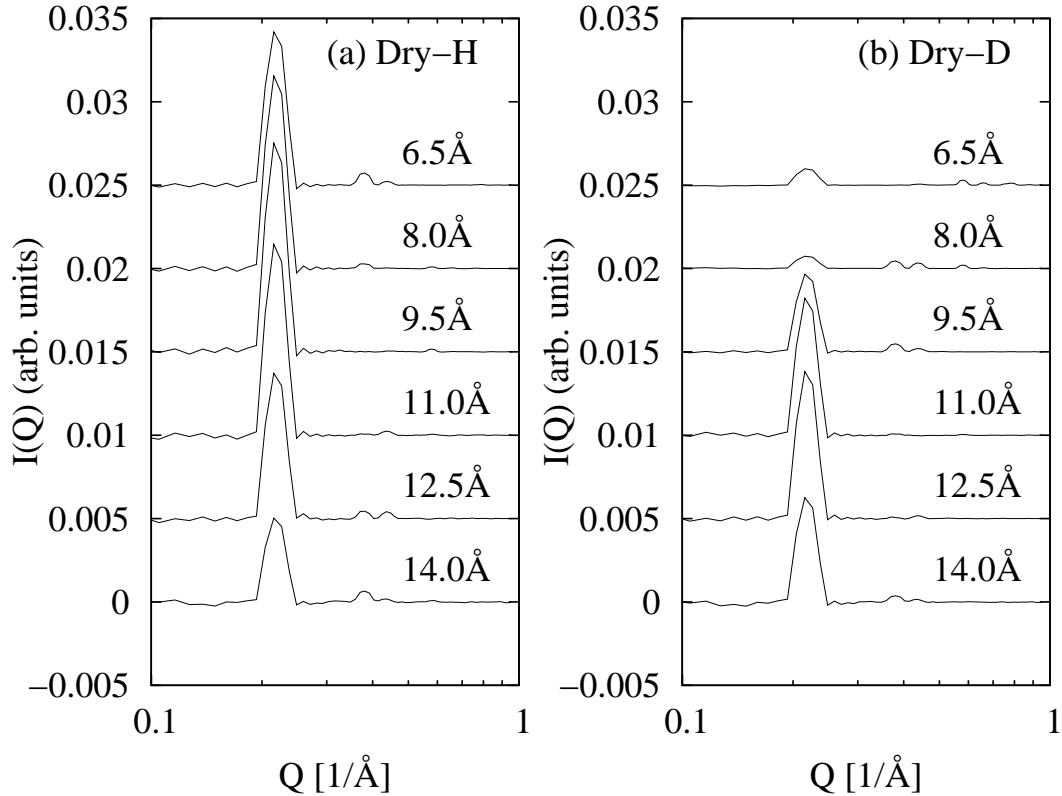}
\caption{Calculated Bragg intensity profiles as a function of pore radius ($r_P$) , assuming a silanol fraction of 0.2 and an overlap layer thickness of 3.0\AA, for dry-H (a) and dry-D (b) MCM-41-S15 silica. The different pore radii are shown on the individual curves which are shifted vertically for clarity.}
\label{fig-iofqradius} 
\end{figure}

In fact there is another reason why, for the samples of Mancinelli \textit{et al.} \cite{manc2009:1}, the pore radius may be of order 12.5\AA. This is to do with the relative intensities of the (100), (110) and (200) Bragg peaks for the dry substrate. Figure \ref{fig-iofqradius} shows the simulated scattering pattern for MCM-41-S15 with the pores empty for a series of pore radii encompassing the values given in Figure \ref{fig-massadsorbed}. Several factors compel us to insist that the correct pore radius for this material is near 12.5\AA. 

Firstly we can see that for smaller pore radii, the (100) peak in the dry-D sample is much weaker than for the dry-H sample: this is quite different from the behaviour seen in the measured data, Figure \ref{fig-mdcs}, where the (100) peak has almost the same amplitude for these two samples. Secondly in the data the intensity of (110) and (200) is significantly lower for the dry-D sample compared to the dry-H sample, with the (110) weaker than (200). This situation is only found in the simulation for a pore radius of 12.5\AA. Thirdly for pore radii smaller than 11.0\AA\ higher order reflections beyond (200) are found in the simulation for the dry-D sample which are not present in the diffraction data. Finally for a pore radius of 14.0\AA\ the (100) peak has markedly higher amplitudes for dry-D compared to dry-H. This trend is not found in the diffraction data, and the ratio of (100) to (110) and (200) intensities is much smaller than for smaller pore radii, suggesting again that a pore radius as large as 14.0\AA\ is incorrect.

\begin{figure}[htp]
\centering
\includegraphics[width=1.0\textwidth]{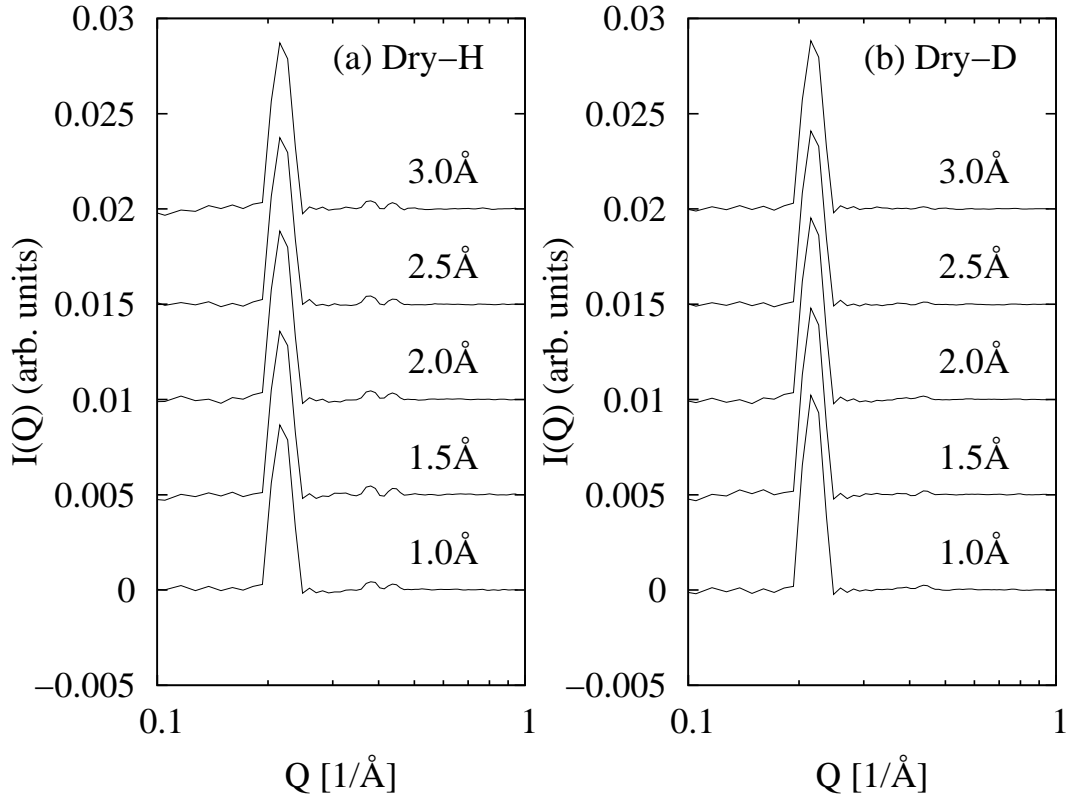}
\caption{Calculated Bragg intensity profiles as a function of overlap layer thickness $(r_P-r_I)$ for a pore radius ($r_P$) of 12.5\AA, assuming a silanol fraction of 0.2, for dry-H (a) and dry-D (b) MCM-41-S15 silica. The overlap layer thicknesses are shown on the individual curves, which are shifted vertically for clarity.}
\label{fig-iofqtshelf} 
\end{figure}

The effect of changing the thickness of the overlap region, $(r_P-r_I)$, using a fixed pore radius, $r_P=12.5$\AA\ and silanol fraction (0.2), is shown in Fig. \ref{fig-iofqtshelf}. It is seen that in fact the sequence of Bragg peak intensities is rather insensitive to the overlap thickness at this pore diameter, although it will also be seen that for an overlap thickness less than 3.0\AA, the (100) peak in the dry-D sample becomes progressively larger than the corresponding peak in the dry-H sample: this is the opposite of the experiment, where this peak is almost the same intensity in both samples. An equivalent graph (not shown here) was obtained when the pore radius was set to 8.0\AA, where the Bragg peak intensities were even more invariant to the overlap thickness. In that case however the ratio of (100) intensities for dry-H and dry-D samples is completely wrong compared to experiment as already shown in Fig. \ref{fig-iofqradius}.

\begin{figure}[htp]
\centering
\includegraphics[width=1.0\textwidth]{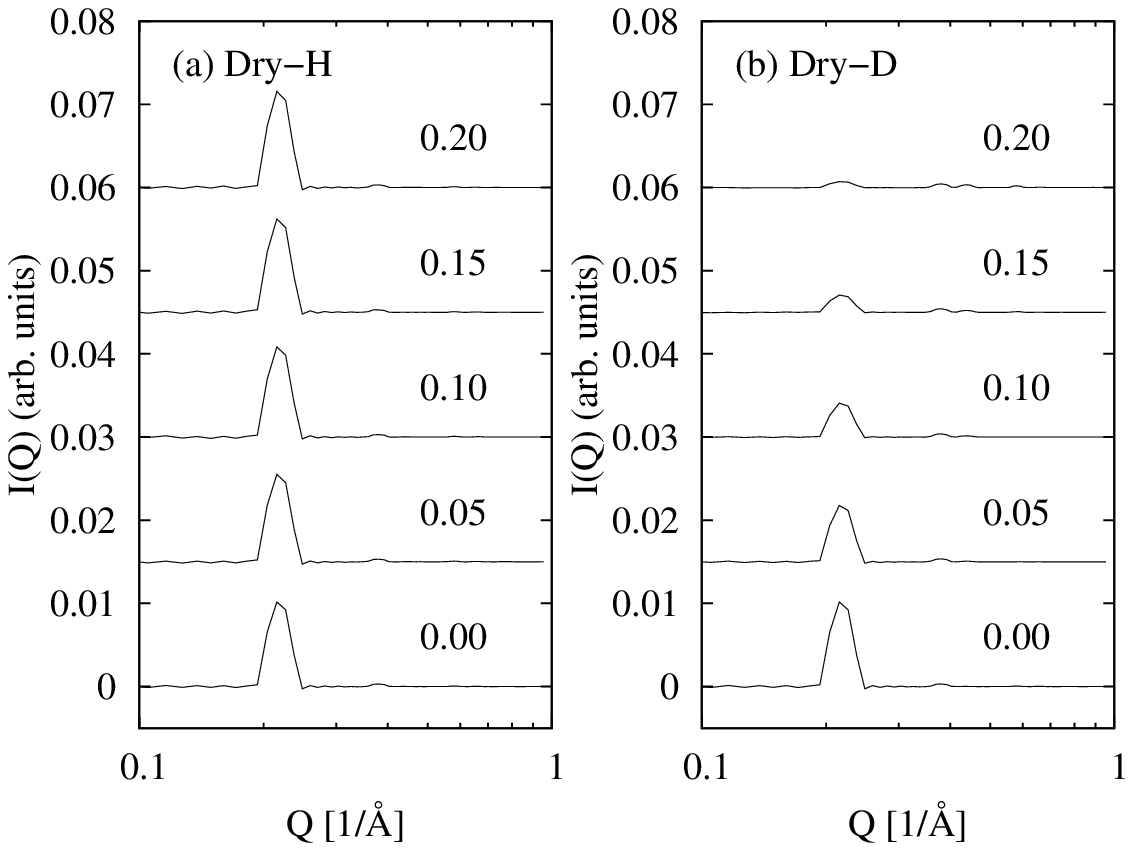}
\caption{Calculated Bragg intensity profiles as a function of silanol fraction $x$ for a pore radius ($r_P$) of 8\AA\ and overlap thickness of 3\AA\ for dry-H (a) and dry-D (b) MCM-41-S15 silica. The silanol fractions are shown on the individual curves, which are shifted vertically for clarity.}
\label{fig-iofqsfrac} 
\end{figure}

For a pore radius of 12.5\AA\ the effect of changing the silanol fraction at constant overlap thickness was negligible and so is not shown here, but for a pore radius of 8\AA\ the effect is quite marked, as shown in Fig. \ref{fig-iofqsfrac}. It will be noted that the (100) dry-H and dry-D Bragg peak intensities are only the same for this pore radius (as in the experiment) when the silanol fraction is set to zero. Since silanol groups are essential for water absorption in MCM-41 this seems like an unlikely scenario.  

The above analysis highlights the sensitivity of the Bragg peak intensities to the choice of pore radius, overlap thickness and silanol fraction. Combining the evidence of amount of water adsorbed per mass of substrate with this diffraction evidence, it seems pretty certain that the pore radius in the present sample of MCM-41-S15 was close to 12.5$\pm$0.5\AA. This is in contradiction to previously stated values for this radius. Traditionally estimating the pore radius in these porous silicas has been a matter of some uncertainty.  \cite{MCM412001:1}. Almost certainly this arises from the variety of methods that are invoked to estimate this radius, which usually involve measuring the absorption isotherms of different gases adsorbed by the surface and using this information to estimate the surface area per unit length of the pore. This amount will depend intrinsically on the nature of the gas-surface or liquid-surface interaction, which can vary markedly from material to material. On the other hand the present diffraction evidence from the dry substrate, relies on no such assumptions, and so should be more reliable. The same diffraction method could be used to estimate the average pore radius in other porous materials where the pores are arranged in a regular lattice.

\subsection{\label{wetdensityprofile}Density profile in wet MCM41}

Using only the diffraction evidence from the three Bragg peaks, (100), (110) and (200), probably it would therefore be difficult to guarantee a unique solution in determining the density profile. In this case however the earlier empirical potential structure refinement (EPSR) simulations can be used as a guide \cite{manc2009:1, manc2010:2}. In those simulations the water density at 300K in the interfacial region was observed to be $\approx$2 times larger than the water density in the core region, with a decaying density in the overlap region. On the other hand at 210K the ratio of interfacial to core regions densities dropped to $\approx$1.5 under the constraint that the overall water density in the pore was not allowed to change. Are these variations consistent with the observed changes in the Bragg peaks between 300K and 210K?

\begin{figure}[htp]
\centering
\includegraphics[width=1.0\textwidth]{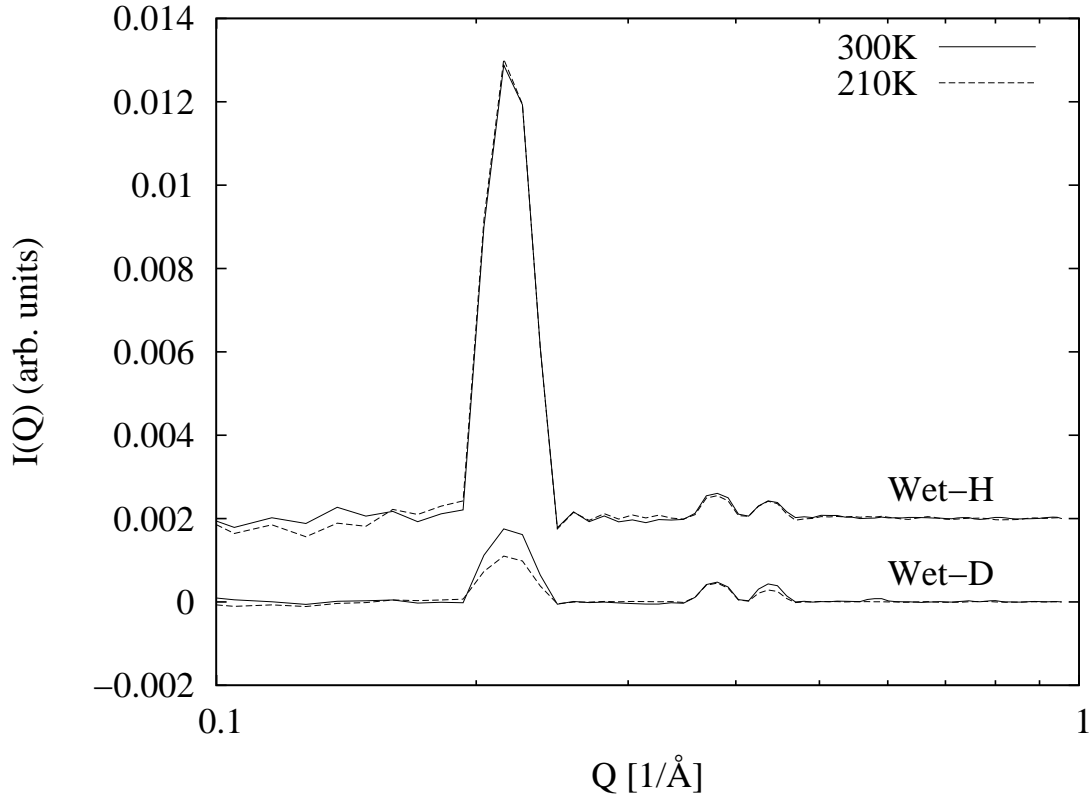}
\caption{Simulated Bragg intensity profiles for  MCM-41-S15 silica assuming a silanol fraction of 0.2, for wet-H and wet-D materials at 300K (solid) and 210K (dashed). The H data are shifted vertically for clarity. Instrumental broadening, which is likely to be $Q$ dependent and significant due to the low scattering angle at which the data have to be measured, as well grain size effects, single atom scattering, and pore size distribution effects, have not been included in these simulations.}
\label{fig-iofqwet} 
\end{figure}

To answer this question the pore radius was set to $r_P=12.5$\AA\ as in the simulations described above, with the core region extending to $r_C=6.5$\AA\ and the interfacial region to $r_I=9.5$\AA. The overlap thickness was therefore 3.0\AA. The silanol fraction was set to 0.2 as above, while the average water density was set to $\langle\rho_W\rangle=0.07$ atoms/\AA$^3$ for both 300K and 210K to give an overall water adsorption of 0.32g H$_2$O/g substrate. Although lower than the stated value in \cite{manc2009:1}, this value is consistent with the observed change in high $Q$ scattering level on going from dry to wet substrate (see Section \ref{experiment}) and gives the correct simulated peak height dependence with temperature as observed in the diffraction data. At 300K the relative water densities were set to $\rho_C=$1, $\rho_I=$2, and $\rho_O=$0.7 for the core, interfacial and overlap regions respectively, with the actual densities chosen to give the specified overall density. At 210K the same densities were set to 1, 1.5, and 0.9 respectively, giving a more uniform density profile than at 300K, as found in \cite{manc2010:2}. Figure \ref{fig-iofqwet} shows the results of these simulations and Table \ref{tab-simrelinten} gives the simulated relative peak heights to be compared with the measured values shown in Table \ref{tab-relinten}. 

\begin{table}
\caption{\label{tab-simrelinten} Relative heights of the simulated (100) Bragg peak for MCM-41-S15 for dry and wet hydrations and at 300K and 210K as a function of hydrogen isotope, H or D, calculated according to the simulations described in this paper. All intensities are normalised to the intensity of the dry-D sample at each temperature.\hspace*{12pt}}
\begin{center}
\begin{tabular}{| l | c | c |}
\hline
Sample & 300K & 210K \\
\hline
Dry-D  & 1.00 & 1.00 \\
Dry-H  & 0.98 & 0.99 \\
Wet-D  & 0.20 & 0.12 \\
Wet-H  & 1.25 & 1.24 \\
\hline
\end{tabular}
\end{center}
\end{table}

\begin{figure}[htp]
\centering
\includegraphics[width=1.0\textwidth]{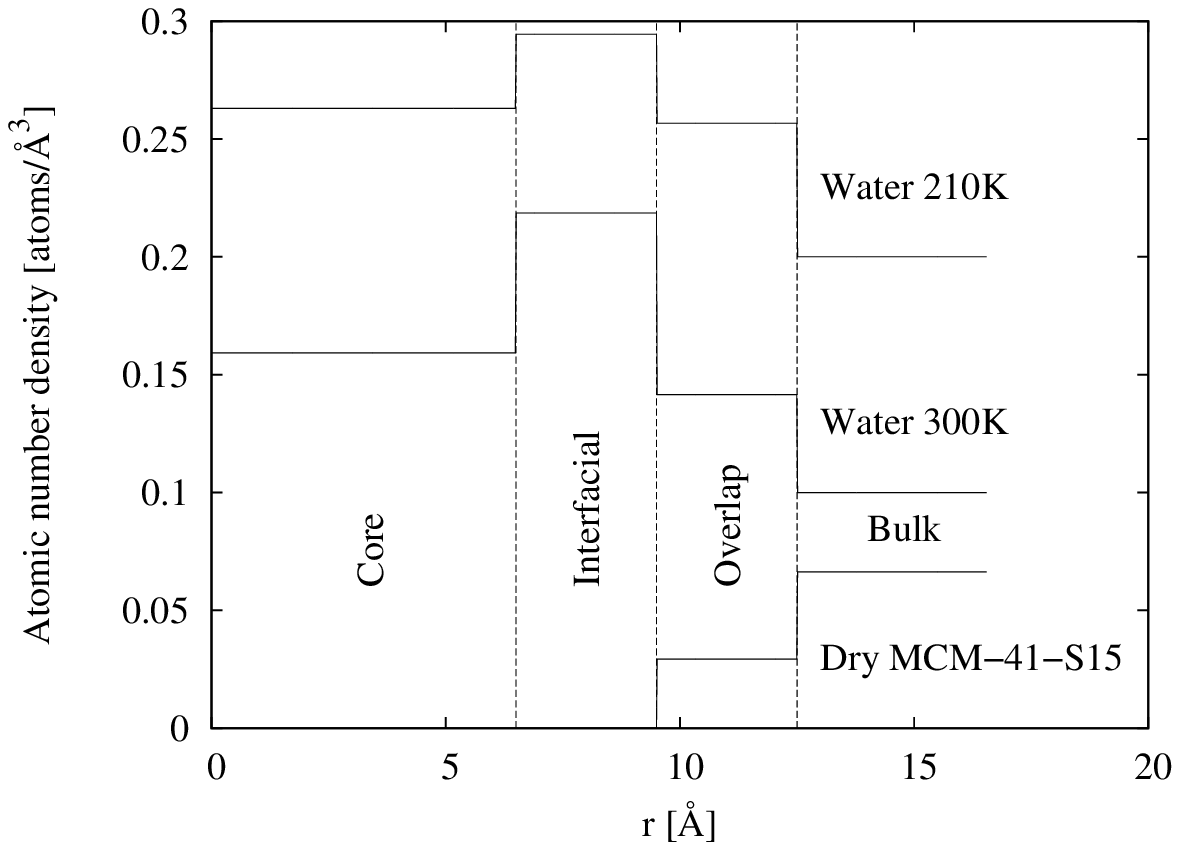}
\caption{Simulated atomic number density profile for dry MCM-41-S15 (bottom), for confined water at 300K (middle, shifted upwards by 0.1 atoms/\AA$^3$) and confined water at 210K (top,, shifted upwards by 0.2 atoms/\AA$^3$). Vertical dashed lines demark the core, interfacial, overlap and bulk regions of the pore.}
\label{fig-densityprofiles} 
\end{figure}

Clearly the qualitative trend of the diffraction data with temperature and between H and D samples is reproduced by these simulations. The main discrepancies are that the ratio of wet over dry intensity for the (100) peak in the simulation of the H sample is about 1.25, instead of 1.4 as observed in the experiment (see Table \ref{tab-relinten}), and for the wet D samples the heights of the (110) and (200) peaks are probably too large compared to the experiment. Hence some refinement of these values might be possible to improve the agreement with experiment. Figure \ref{fig-densityprofiles} shows the subtrate density profile and the water density profiles at 300K and 210K used in this simulation.

\section{\label{conclusion}Discussion and conclusion}

At this stage an exhaustive study of all possible values of the density profile that might be consistent with the Bragg diffraction data has not been attempted. In addition only limited refinement of the values used in Section \ref{wetdensityprofile} has been performed, although such refinement should be straightforward. What does come out very clearly in this analysis however is that the height of the Bragg peaks, particularly those arising from the deuterated samples are highly sensitive to the choice of pore radius, overlap thickness, silanol fraction, and density profile across the pore, and so therefore cannot be used to state conclusively what is the density of water confined in porous silica, since a change in the density profile, such as that shown in Figure \ref{fig-densityprofiles} can also have a marked effect on Bragg intensities. In particular the density profiles proposed in the earlier EPSR studies \cite{manc2009:1, manc2010:2} appear to be consistent with the observed Bragg peak intensities, within the approximations of the present work, which does not include an atomistic simulation as previously used. There is a difference between the present work and the previous work in that the assumed pore radius is here much larger (12.5\AA) than assumed previously (9.6\AA). However in the earlier work there was significant penetration of the water into the silica matrix, which is covered by the overlap region in the present model, so in practice the two models may not be so different. Whatever the correct values for these parameters it is clear much more work needs to be done characterising the pore before any conclusions can be drawn about the nature of the water absorbed within it.

\begin{figure}[htp]
\centering
\includegraphics[width=1.0\textwidth]{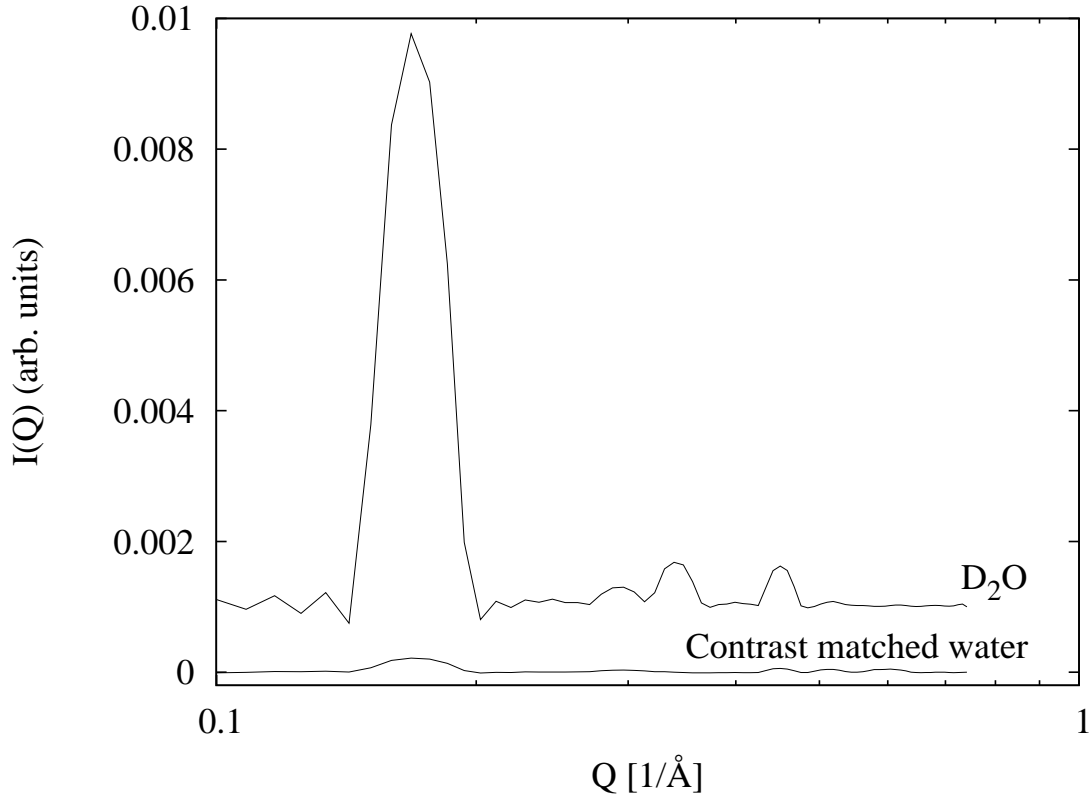}
\caption{Simulated Bragg intensity profiles for  MCM-41 silica with the same pore separation (42.7\AA) as given in \cite{langmuir2002:1} assuming a pore radius of 12.5\AA, an overlap thickness of 3\AA, and silanol fraction of 0.2, for contrast matched water (bottom, H:D ratio 0.4:0.6) and D$_2$O (top). The latter curve is shifted vertically for clarity. The density profiles across the pore and used in this calculation are taken from Fig. \ref{fig-densityprofiles} at 300K. }
\label{fig-iofqnull} 
\end{figure}

Another feature of the literature on confined water is the (often) implicit assumption that the density profile in these systems is uniform across the pore. The highly charged nature of the silica surface, the fact that the surface is usually disordered, and the method of formation of these materials from cylindrical micelles, make this statement implausible at the atomistic level. Nonetheless evidence for this assertion comes from experiments in which the ratio of H$_2$O:D$_2$O is chosen so that the scattering length density of the so-called contrast matched water is the same as the underlying silica matrix. Assuming the substrate is pure SiO$_2$ this ratio is approximately 0.4H$_2$O:0.6D$_2$O, and it is observed that within statistical accuracy the Bragg peaks disappear under these conditions \cite{langmuir2002:1}. It will be noted however in that work, that the (100) Bragg peak occurs at 0.17\AA$^{-1}$, which would imply a pore separation of 42.7\AA\ - significantly larger than any values considered previously in this paper. In Fig. \ref{fig-iofqnull} the simulated Bragg intensities for the same contrast matched water using a pore separation of 42.7\AA\ are compared with those from a simulation of wet-D MCM silica, using the same pore size (12.5\AA), overlap thickness (3\AA) and silanol fraction (0.2) as in Fig. \ref{fig-iofqwet}. It can be seen that although the simulated intensities do not go identically to zero, the Bragg peaks drop dramatically in amplitude and might be statistically too small to be observable. This occurs despite there being a highly non-uniform simulated density profile in the pore in this case, as in Fig. \ref{fig-densityprofiles}. Indeed the authors of this earlier work state that as water is desorbed from the pore, the density profile becomes non-uniform, with the highest density occurring near the edge of the pore, as in the present case. Since the previous experiments on water in confined geometry \cite{manc2009:1} were performed below saturation conditions, the finding of a non-uniform density profile in that case is perfectly reasonable. Here I have shown that negligible Bragg intensity with contrast matched water cannot be used to claim the density profile is uniform.

A serious discrepancy with the earlier work of Liu \textit{et al.} \cite{savedrecs-32007:1} is the pore radius, which as discussed in Section \ref{poreradius} is significantly different here to that stated in \cite{savedrecs-32007:1} for ostensibly the same material. The pore diameter determined from the present diffraction data, 25\AA, is near the nominal limit of pore diameters at which water will show a feature in the differential scanning calorimetry trace corresponding to freezing on supercooling, \cite{findenegg2008:1}. Nonetheless close inspection of the wet-D sample at 210K in the region of the main water diffraction peak near $Q = 1.8$\AA$^{-1}$ (see Figure \ref{fig-mdcs}, inset) reveals three small peaks which appear closely analogous to those found in crystalline ice \cite{manc2010:1}, and other relatively sharp peaks appear at higher $Q$ values in the 210K data which are not present in the 300K data. (See also Figure 2 of \cite{manc2010:2}.)  Indeed the shift of the main diffraction peak in D$_2$O from $Q\approx 2.0$\AA$^{-1}$ at 300K to $Q\approx 1.8$\AA$^{-1}$ at 210K is similar to what is observed in heavy water when it freezes. Given the pore size broadening effect described by \cite{manc2010:1}, these peaks would be substantially broadened for a pore diameter of 25\AA. If the same crystal-like ordering transition occurred in the materials used to study the dynamic cross-over in supercooled water, then that cross-over would have a completely different explanation to the one normally advanced. This work of course does not rule out the possible dynamic crossover scenario to explain the apparent fragile to strong behaviour on cooling water absorbed in MCM-41 (or other substrate materials), but it does raise serious questions about the interpretation of existing experimental data such as that listed in the Introduction. The necessary pore size and wider $Q$ atomistic information on the samples used to absorb the water is often unavailable: without that information the claims for observation of a FS transition or true density minimum are unsubstantiated, at least in the present case of MCM-41. Similar characterisation studies will needed for other substrates before similar conclusions can be drawn. Discrepancies to emerge in the present work on MCM-41 are that the pore separation clearly varies markedly from sample to sample, and there are significant doubts about what is the correct pore diameter: these parameters will have a profound effect on the amount of water that can be absorbed by the substrate.

Quantitatively the density profiles presented here in Figure \ref{fig-densityprofiles} are not identical to those obtained in the recent EPSR simulations, \cite{manc2010:2}, but have considerable similarities. The main differences are in the overlap region where more water appears to penetrate the substrate in the present simulations than was found previously. In particular this overlap water increases on cooling to 210K whereas in the EPSR simulations it apparently decreases. Whether this difference can be resolved by formulating a version of EPSR that specifically fits the low $Q$ Bragg peaks as well as the wider $Q$ remains to be seen, but qualitatively at least it is correct to say the previously simulated density profiles are consistent with the observed change in Bragg peak intensity with hydration and hydrogen isotope.

The present results of course do not categorically rule out the possibility of there being a density decrease on supercooling water in MCM41 silica. It is now clear however that the observed trends in Bragg peak intensities with temperature can be understood in terms either of a density change, or a change in density profile, or a combination of these two scenarios. What the present analysis demonstrates is that observation of a change in Bragg peak intensity with temperature does not, by itself, provide conclusive evidence for a change in density. It could equally be explained as a change to a more homogeneous density profile on lowering the temperature, and this latter scenario is consistent with previous atomistic simulations of the wider $Q$ data at several temperature values. Hence the evidence for a density change is at best ambiguous at this point in time. A corollary of the density decreasing with cooling scenario is that if the density decreases, then the excess water will need to be excluded from the pore. This water can then freeze when it is eliminated, but will be perfectly visible in neutron scattering or other dynamics experiments. Such frozen water could also modify Bragg peak intensities, but more importantly might give a false signal of a dynamical transition when it is actually a freezing transition.

More generally, for determining the scattering characteristics of both wet and dry ordered porous silicas, it appears that the present approach of simulating the Bragg peak intensity profile can provide invaluable information about both the pore size and the distribution of atomic density within it. This information is essential if we are to make accurate assessments of the nature of dynamic and structural transitions in the pore fluid.

\ack I am indebted to Rosaria Mancinelli, Fabio Bruni and Maria Antonietta Ricci for access to  the original data for this experiment, and for numerous discussions about the methods and content in this paper.

\section*{References}
\bibliography{References}

\end{document}